\documentclass[reprint,superscriptaddress,longbibliography,
 amsmath,amssymb,
 aps,prb
]{revtex4-1}

\usepackage{graphicx}
\usepackage{dcolumn}
\usepackage{bm}

\begin{document}

\preprint{APS/123-QED}

\title{Bulk-Edge Correspondence in Fractional Quantum Hall States}

\author{Bin Yan}
\affiliation{
 Department of Physics and Astronomy, Purdue University, West Lafayette
}
\author{Rudro R. Biswas}
\affiliation{
 Department of Physics and Astronomy, Purdue University, West Lafayette
}
\affiliation{
Purdue Quantum Center, Purdue University, West Lafayette
}
\author{Chris H. Greene}
\affiliation{
 Department of Physics and Astronomy, Purdue University, West Lafayette
}
\affiliation{
Purdue Quantum Center, Purdue University, West Lafayette
}

\date{\today}

\begin{abstract}
We substantiate a complete picture of the ``bulk-edge correspondence" conjecture for topological phases. By studying the eigenstates in the entanglement spectrum for both the ideal and realistic Coulomb ground states of the fractional quantum Hall system, it is verified that the eigenstates in the universal part of the entanglement spectrum purely lie in the Hilbert space of the edge excitations projected onto the physical Hilbert space of the subsystem itself. Hence, not only are the eigenlevels in the entanglement spectrum in one-to-one correspondence with the eigenenergies of an effective dynamical edge Hamiltonian, but all the eigenstates are confirmed to be the actual (projected) edge excitations of the subsystem. This result also reveals the possibility of extracting the full information of the edge excitations from the state of the subsystem reduced from a geometric cut of the pure ground state of the total system in topological phases.
\end{abstract}

\maketitle

\section{Introduction}
Since the discovery of the fractional quantum Hall effect \cite{Tsui:1982kq,Girvin:1987}, a new paradigm of phases of matter which is of a topological nature \cite{Wen:1989cy,Wen:1990fd,Nayak:2008dp} and yet does not fit into the conventional symmetry-breaking picture comes into play. These novel phases of matter exhibit spectacular behaviors, such as fractional charges \cite{Laughlin:1983hk}, (non-)Abelian anyons \cite{Leinaas:2007iv,*Wilczek:1982cm,*Wilczek:1982be}, or topologically protected gapless edge excitations. In absence of local order parameters, these phases are usually characterized by certain global properties such as the ground state degeneracy \cite{Wen:1990fd}. Despite intense effort, a complete understanding of such topological ordered phases still remains elusive.

Among various approaches to characterize these intriguing zero-temperature phases of matters, entanglement plays an increasingly crucial role \cite{Amico:2008en}. It is believed that ground states of topological ordered systems are able to build up long range entanglement \cite{Chen:2010gb}. One way to study the entanglement in a many-body system is to look at the correlation between a subsystem and its complement in the total system. The entanglement entropy, i.e., the subsystem von Neumann entropy, captures the entanglement between the subsystem and the rest of the total system. In typical ground states of many-body systems, the entanglement entropy follows an area scaling law \cite{Eisert:2010hq}. For a state in a topological phase, besides the usual term that is proportional to boundary area, there emerges a constant sub-leading term that is characteristic of the topological nature of the system. This quantity, called topological entanglement entropy \cite{Hamma:2005je,Levin:2006ij,Kitaev:2006dn}, is one of the most recognized signatures of states with topological phases.

In subsequent work \cite{Li:2008cg}, Li and Haldane proposed and numerically substantiated that more information about the underlying topological state can be obtained by studying instead the full spectrum of the reduced density matrix, $\rho_A$, of a subsystem A. Writing $\rho_A=e^{-H_A}$, the entanglement spectrum is defined as the spectrum of the effective Hamiltonian $H_A$. The eigenvalues $\xi_i$ and eigenstates $|\Psi\rangle^A_i$ of $H_A$ can be alternatively extracted from the Schmit decomposition of the total state, 
\begin{equation}
|\Psi\rangle=\sum_i e^{-\xi_i/2} |\Psi\rangle^A_i \otimes |\Psi\rangle^B_i.
\end{equation}
Though $H_A$, in general, is not a real dynamical Hamiltonian, it is ultimately related to the physical edge of the subsystem: In the quantum Hall system, the number of the low-lying universal eigenenergies, which are separated by a finite gap \cite{Li:2008cg,Thomale:2010it} from other generic levels, are in one-to-one correspondence with the edge modes describable by a macroscopic edge theory.%, at least for small angular momentum excitations.

Both the entanglement entropy and the counting structure of the entanglement spectrum can be naturally interpreted if a relation between the bulk and the physical edge of the subsystem is established. This relation, termed the bulk-edge correspondence \cite{Qi:2012jv,Swingle:2012kj,Dubail:2012kx}, states that the entanglement Hamiltonian corresponds to an effective dynamical local Hamiltonian acting on a 1-D edge system. The bulk-edge correspondence, originally sketched in the early work of entanglement entropy \cite{Kitaev:2006dn}, has been extensively studied in various fields \cite{Liu:2013fm,Santos:2013bf,Ho:2015fh,Chen:2016ks,Ho:2017cza,KochJanusz:2017cv}. This correspondence emerged in part from general arguments based purely on the topological properties of the system and on the standard renormalization-group method \cite{Qi:2012jv}, in part from observations of geometric aspects and the Lorentz invariance of the emergent effective theory \cite{Swingle:2012kj}, and in part through study of
model wavefunctions specific to the quantum Hall system constructed from conformal blocks \cite{Dubail:2012kx}.
These arguments are only valid provided certain assumptions hold in the thermodynamic limit. 

\begin{figure}
\includegraphics[width=8.5cm]{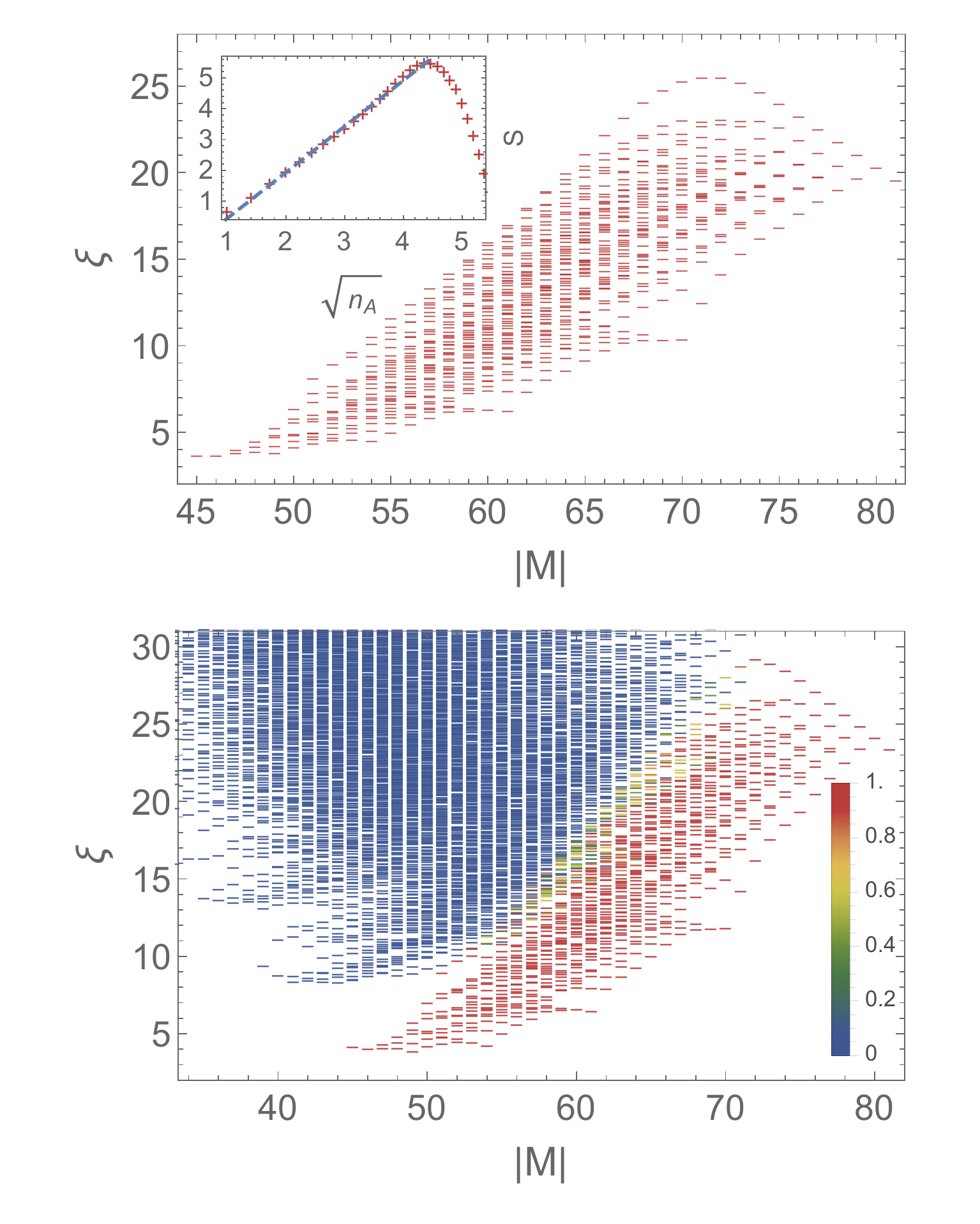}
\caption{\label{fig:spectrum}
Entanglement spectrum of the 17-orbital subsystem at the 6-particle sector. The total system of 12 particles in 34 total orbitals is prepared in (Upper) the ideal Laughlin state and (Lower) the realistic Coulomb interaction ground state. The color of the pseudo-energy levels indicate the overlaps between the corresponding entanglement spectrum eigenstates and the edge states of the subsystem. Inset: Entanglement entropy v.s. the square root of the number of orbitals ($n_A$) in subsystem $A$, which is proportional to the length of the boundary. The blue dashed line is our fit to the first 17 points, which are in the linear region away from the edge of the total system.} %The extracted topological entanglement entropy is $1.040$, close to the theoretical value ln$3$.
\end{figure}

In this work, we argue that a complete picture of the bulk-edge correspondence should consist of two pieces. One is that the eigenvalues $\xi_i$ are in one to one correspondence with the eigenvalues of an effective edge Hamiltonian. The other is that the universal eigenstates $|\Psi^A_i\rangle$ that appear in the bulk density matrix (See Fig. \ref{fig:spectrum} for an illustration of the universal levels) are all real edge states of the subsystem. To be precise, the universal levels are these appearing in the entanglement spectrum of the ideal model wavefunctions, as well as the part of the spectrum having similar structures for the case of generic interactions. By comparing the entanglement spectrum eigenstates with the actual edge excitations, we are able to identify the universal branch of the spectrum (See Fig. \ref{fig:spectrum} for an example). For the first part of the bulk-edge correspondence in the quantum Hall system, the only direct numerical evidence was provided in Ref. \cite{Dubail:2012kx}, where the spectrum of a perturbative local Hamiltonian of an edge system described by a conformal field theory was computed and shown to match quite well the entanglement spectrum obtained for a real space partition of the total system. 
To the best of our knowledge, a direct verification of the latter part of the correspondence is still missing. Since an effective edge theory usually describes the degrees of freedom that are distinct from those in the original bulk system, direct comparison of their eigenstates is much harder than comparing their eigenenergies. Thus, typical studies of the entanglement spectrum usually focus on the counting structure of the eigenlevels. In the present work, by analyzing the detailed information of the entanglement spectrum eigenstates, we complete the verification of the missing piece of the bulk-edge correspondence for both the model wavefunction and the realistic ground state of Coulomb interactions.

\section{Entanglement Spectrum and Edge Physics}
The system studied here consists of a few spinless fermions in the lowest Landau level on a two-dimensional disk.
In this work, system partitions are restricted to the subsets of the single particle magnetic orbitals in the symmetric gauge (with unnormalized analytic wavefunction $z^{|m|} e^{-|z|^2/4}$, where $z=x-iy$ is the two dimensional complex coordinate and $m=0,-1,-2,...$ is the angular momentum quantum number). In principle, orbital cut is a parition in momentum space and is fundamentally nonlocal. However, in the lowest Landau level, these single particle orbitals have localized ring shapes with narrow width in the order of the magnetic length. Thus in many cases the orbital partition mimics approximately the real space partition and provide valuable information of the system. It is also worthwhile to mention that another commonly used scheme, based on a cut of particle number \cite{Haque:2007il,Zozulya:2007jw,Sterdyniak:2011cp}, has usually been adopted for the study of quasi-hole excitations. In some particular model states, the level counting in the orbital entanglement spectrum and the one in the particle entanglement spectrum, which correspond to quasi-hole excitations in the bulk, were shown \cite{Chandran:2011ex} to be ultimately connected. This relation, also termed ``bulk-edge correspondence", is not to be confused with the one we are considering in the present work.

In the subsystem, the particle number and total angular momentum are still good quantum numbers; Thus, the reduced density matrix of a subsystem contains distinct sectors which do not couple with each other. As an illustration, Fig. \ref{fig:spectrum} compares the standard entanglement spectrum for the ideal Laughlin state with that of the realistic Coulomb ground state of 12 particles in the lowest Landau level. For entanglement spectra of the quantum Hall states in other geometry and partition schemes, see also Refs. \cite{Zozulya:2007jw,Lauchli:2010jw,Sterdyniak:2011cp,Dubail:2012dn,Sterdyniak:2012fl,Rodriguez:2012bl}. The subsystem studied in this work involves the inner most $N/2$-orbitals (with N the even number of total orbitals), which is the largest system that is sufficiently far away from the edge of the total system, in the sense that it is still in the linear region of the entanglement entropy (See the inset in Fig. \ref{fig:spectrum}). In this region, edge effects are supposed to be removed as much as possible. By analyzing the fraction of each eigenstate that resides in the Hilbert space of edge modes (by the procedure explained in the following), we are able to identify all the universal levels apart from the generic levels in the case of Coulomb interactions. It is observed that the universal and generic parts barely mix with each other, albeit at larger angular momentums they start to mix, in contrast to the low angular momenta, where there is a finite gap between them.

To prove the bulk-edge correspondence, on the ``edge'' part, we use the well-known series of model wavefunctions \cite{Wen:1992tb} for the (neutral) edge excitations of the Laughlin ground state. These analytic model wavefunctions are generated by the power sum symmetric polynomials $s_n=\sum_i z_i^n$, $n=1,2,3...$. To build an edge mode with angular momentum $\Delta M$ (this is the additional angular momentum carried by the edge mode beyond the ground state angular momentum), one generates an integer partition $|\Delta M|=\sum_i n_i$, where $n_i$ are positive integers. The corresponding (unnormalized) edge state is constructed as the product of the Laughlin wavefunction and a symmetric polynomial:
\begin{equation}\label{eq:edge}
\Psi_{edge}=\prod_i s_{n_i} \prod_{i<j} (z_i-z_j)^{1/\nu} \prod_i e^{-\frac{|z_i|^2}{4}},
\end{equation}
where $\nu$ is the filling factor. In general $1/\nu$ takes arbitrary odd integer values, while in the following numerical study we focus on $\nu=1/3$ only. For a given $\Delta M$, the number of different modes is thus the number of integer partitions of $|\Delta M|$. These states are the only zero energy states for the case of the Haldane $V_1$ pseudo-potential \cite{haldane:1983} in the lowest Landau level, and they describe the gapless edge excitations of the Laughlin ground state. In the thermodynamic limit, the Hilbert space of the edge modes generated by the order-$n$ power sum polynomials is identical to that generated by the U(1) Kac-Moody algebra \cite{Wen:1990ele,Wen:1990chi,Wen:1992tb} in a macroscopic theory of the edge physics. However, for finite size systems, when the orders of the symmetric polynomials become larger than the number of particles, these states are no longer linearly independent. In fact, the number of linearly independent symmetric polynomials of order $n$ is the number of partitions of $n$ into at most $N$ parts, where $N$ is the number of particles and the number of variables in the polynomial. Denote this restricted partition as $P(n,N)$. For each such integer partition $n=\sum_i^N n_i$, one alternative way to construct the edge state is to use the corresponding order-$n$ monomial symmetric polynomial $s_n=\sum_{p}(\prod_i z_i^{n_i})$, where the summation is taken over all permutations of $z_i's$ such that the polynomial is symmetrized. The leads to a correction of the number of edge modes for finite size systems compared with the macroscopic theory for systems in the thermodynamical limit.

Another finite size effect arises when comparing the edge modes with the eigenstates in the entanglement spectrum. That is, the ideal edge modes might involve single particle orbitals that are outside the region of the subsystem under consideration. %(Fig.\ref{fig:compare} shows the actual spectra of the subsystem and the edge state spectra with open boundary condition.) 
These additional degrees of freedom need to be traced out in formatting the appropriate Hilbert space of the subsystem \cite{Dubail:2012kx}. Precisely speaking, the Fock space spanned by the free edge states are projected onto a subspace involving only orbitals that are resident in the subsystem. This inevitable projection procedure was also discussed in Ref. \cite{Dubail:2012kx} for the case the real space partitions. In this manner, the projections of the original linearly independent edge modes might become linearly dependent, such that the projected Fock space could have a smaller dimension than the original Fock space of the free edge states. For illustration, Table \ref{table} lists the number of free edge modes, the number of independent projected edge modes, and the number of observed energy levels in the entanglement spectrum for the 5- and 4-particle sectors. The observed level counting in the primary 5-particle sector of the entanglement spectrum matches precisely the reduced dimension of edge modes Hilbert space at every angular momentum sector. In the large particle number limit, the restricted integer partition reduces to the regular integer partition. This is in agreement with the original conjecture for the counting structure of the entanglement spectrum in Ref. \cite{Li:2009tn}. 

\begin{table}[h]
\begin{tabular}
{ c c c c c c c c c c c c c c c c c c  } 
 \hline
  \hline
 $|\Delta M|$ &0&1&2&3&4&5&6&7&8&9&10&11&12&13&14&15&...\\ 
 \hline
 $P(|\Delta M|,5)$ &1&1&2&3&5&7&10&13&18&23&30&37&47&57&70&84&...\\ 
 \hline
 $D_{p}$&1&1&2&3&5&7&9&11&14&16&18&19&20&20&19&18&... \\ 
  \hline
 $D_{m}$ &1&1&2&3&5&7&9&11&14&16&18&19&20&20&19&18&... \\ 
 \hline
 \\
 \hline
 \hline
 $|\Delta M|$ &0&1&2&3&4&5&6&7&8&9&10&11&12&13&14&15&...\\ 
 \hline
 $P(|\Delta M|,4)$ &1&1&2&3&5&6&9&11&15&18&23&27&34&39&47&54&...\\ 
 \hline
 $D_{p}$ &1&1&2&3&5&6&9&11&14&16&19&20&23&23&24&23&... \\ 
  \hline
 $D_{m}$ &1&1&2&3&4&5&7&7&8&8&8&7&7&5&4&3&... \\ 
 \hline
\end{tabular}
\caption{\label{table} The number of edge modes at various angular momentum $\Delta M$ in the 14-orbital subsystem. The total system has 10 particles occupying 28 orbitals. Upper and lower tables correspond to five and four particles, respectively. $P$ is the number of free edge modes, i.e., the number of restricted integer partitions. $D_p$ represents the dimension of the Hilbert space of projected edge modes; $D_m$ labels the observed multiplicity of levels in the entanglement spectrum.}
\end{table}

\begin{figure}
\includegraphics[width=8cm]{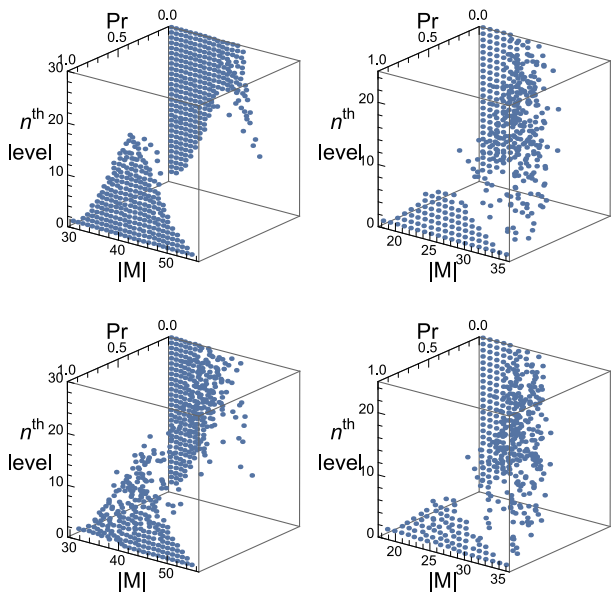}
\caption{\label{fig:overlapf}
The projection probabilities for the whole entanglement spectrum of the 5-particle sector (left) and the 4-particle sector (right) for the case of Coulomb ground state. The cases of both ideal edge states (upper) and realistic edge states of Coulomb interaction are examined.}
\end{figure}

\begin{figure}
\includegraphics[width=8.5cm]{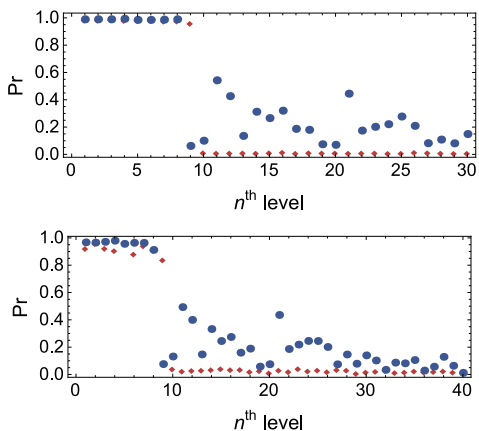}
\caption{\label{fig:overlaps}
Typical patterns for the projection probabilities. Red squares represent the probabilities for 5 particles at $\Delta M=6$ ($|M|=36$). Blue dots correspond to the 4-particle sector at $\Delta M=8$ ($|M|=26$). Projection probabilities are computed using (a) the model edge states for the Laughlin case and (b) the real edge states for Coulomb interactions.}
\end{figure}

To prove that the universal part of the entanglement spectrum is indeed spanned by the edge modes, we compute the probability for each eigenstate in the entanglement spectrum to reside in this space. For a given eigenstate $\psi$ of the entanglement spectrum, the projection probability is defined as

\begin{equation}\label{eq:prob}
Pr(\psi)=|\hat{P}\psi|^2,
\end{equation}
where $\hat{P}$ is the projection operator to the Hilbert space of projected edge modes.

\section{Numerical Results}
We are able to compute the entanglement spectrum for both ideal Laughlin state and Coulomb ground state, and the correponding projection probability defined in previous section, for up to 12-particles using exat diognalization within the lowest Landau level. The data are presented in Fig. 1, which confirms that the eigenstates in the universal entanglement spectrum clearly stand out with a near unity overlap with subsystem edge modes. In the following we illustrate more details with the 10-particle system.

For the ideal Laughlin wavefunction, it is confirmed that all the eigenstates in the entanglement spectrum have unity projection probabilities in the Hilbert space of the projected edge states. 

For the case of Coulomb ground state, both the ideal edge states and the real edge modes of the subsystem are used to analyze the projection probability Eq.(\ref{eq:prob}). The real edge modes are computed from the Coulomb interaction at the corresponding angular momenta with open boundary condition (without performing single particle orbital cut). In the thermodynamic limit, they are supposed to be low-lying modes describing the gapless edge excitations. However, in few-body calculations, there are usually no clear gaps in the spectrum separating the edge modes and the bulk excitations. In this case, we identify each edge state with the aid of high ($\sim$0.97) projection probabilities onto the Hilbert space of the ideal edge wavefunctions (\ref{eq:edge}). These probabilities scale with the system size in the same way as the overlap between the Laughlin model state and the Coulomb ground state. It is known that the latter overlap does not survive in the thermodynamic limit. However, this is not a problem in this case since the edge states would stand out as gapless modes in the spectrum. For the purpose of the few-body calculations in the present work, high overlaps allow us to singles out the right number of edge states, while all other eigenstates of the Coulomb Hamiltonian have sufficiently small projection probabilities ($\sim 0.01$) onto the Hilbert space of the model edge states. 

Fig. \ref{fig:overlapf} and Fig. \ref{fig:overlaps} show the projection probability patterns. Both the primary 5-particle sector and the sector with particle number offset (4-particle sector in this case) show clear high probability plateaus for the universal levels in the entanglement spectrum. In the 4-particle sector, since the number of universal levels is smaller than the dimension of the projected edge modes (as was shown in Table.\ref{table}), the generic levels also have fluctuating non-zero probabilities, in contrast to 5-particle sector, where all the generic levels have nearly zero projection probabilities. They are nevertheless clearly separated from the universal part. 

In the primary sector, since all universal eigenstates in the entanglement spectrum have nearly identity probabilities in the corresponding Hilbert space of projected edge modes, it is concluded that the information of all edge excitations is hidden in the pure ground state of the total system, which can be extracted from the bulk state of the subsystem. This can be viewed as a complementary effect to a similar result in Ref. \cite{Sterdyniak:2011cp}, where it is observed that the primary sector of the entanglement spectrum under particle partition spans the entire space of quasi-hole excitations of the subsystem.

\begin{figure}
\includegraphics[width=6.5cm]{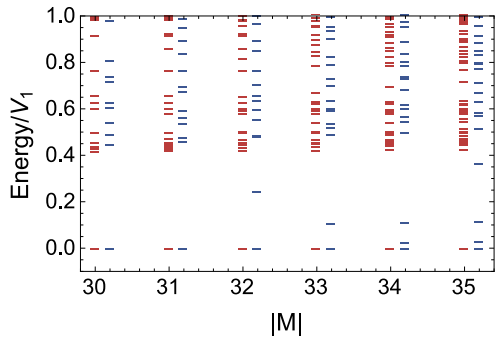}
\caption{\label{fig:compare}
The spectrum of 5 particles under the Haldane $V_1$ pseudo-potential with an open boundary condition (red levels) and with the same boundary condition as the 14-orbital subsystem (blue levels). The blue levels are shifted slightly to the right from their corresponding angular momenta. For the case of an open boundary condition, the degeneracies of the ground states are $1,1,2,3,5,7$, which are identical to the number of the ideal edge states.}
\end{figure}

To further firm up the evidence of the bulk-edge correspondence, we present in Fig. \ref{fig:compare} the actual energy spectrum of 5 particles under Haldane $V_1$ interaction with and without boundary conditions. The degenerate zero-energy states for the latter are model edge states. If we did not use the mechanism of projecting the edge states of the subsystem with open boundaries, the nondegenerate low-lying states of the subsystem alone would have been inadequate to explain all the observed universal levels in the entanglement spectrum (Fig. \ref{fig:spectrum}).

\section{Conclusion}
In conclusion, this work shows that the universal part of the subsystem reduced density matrix, by orbital partition, of the fractional quantum Hall state lies in the Hilbert space of edge modes of the subsystem itself, hence completing the picture of the bulk-edge correspondence conjecture. This result also reveals the possibility of extracting edge excitations from the state of the subsystem reduced by geometric cut of the pure ground state of the total system in topological phases. In the future, it will be informative to apply this method to other systems in topological phases. 
\begin{acknowledgments}
The work of BY and CHG have been supported in part by National Science Foundation Grant NO. PHY-1607180. R.R.B. was supported by Purdue University startup funds and the Purdue Research Foundation.
\end{acknowledgments}

\bibliography{main}
\end{document}